\documentclass[9pt,twocolumn,twoside]{osajnl}
\usepackage{color, soul}
\definecolor{red}{rgb}{1,0,0}
\setulcolor{red}

\graphicspath{{Figures/}}

\journal{ol} % Choose journal (ao, aop, josaa, josab, ol, pr)

\setboolean{shortarticle}{true}
\title{Vortex fiber nulling for exoplanet observations.\\I. Experimental demonstration in monochromatic light}

\author[1,*]{Daniel~Echeverri}
\author[1,$\dag$]{Garreth~Ruane}
\author[1]{Nemanja~Jovanovic}
\author[1,2]{Dimitri~Mawet}
\author[1]{Nicolas~Levraud}

\affil[1]{Department of Astronomy, 
California Institute of Technology, 
1200 E California Blvd, 
Pasadena, CA 91125, USA}
\affil[2]{Jet Propulsion Laboratory, 
California Institute of Technology, 
4800 Oak Grove Dr, 
Pasadena, CA 91109, USA}

\affil[*]{Corresponding author: dechever@caltech.edu}
\affil[$\dag$]{NSF Astronomy and Astrophysics Postdoctoral Fellow}

\ociscodes{(350.1260) Astronomical optics; (300.0300) Spectroscopy; (050.4865) Optical vortices; (060.2430) Fibers, single-mode.}

\begin{abstract}
Vortex fiber nulling is a method for spectroscopically characterizing exoplanets at small angular separations, $\lesssim\lambda/D$, from their host star. The starlight is suppressed by creating an optical vortex in the system point spread function, which prevents the stellar field from coupling into the fundamental mode of a single-mode optical fiber. Light from the planet, on the other hand, couples into the fiber and is routed to a spectrograph. Using a prototype vortex fiber nuller (VFN) designed for monochromatic light, we demonstrate coupling fractions of $6\times10^{-5}$ and $>0.1$ for the star and planet, respectively. 
\end{abstract}

\setboolean{displaycopyright}{true}

\begin{document}

\maketitle
\section{Introduction}

Detecting spectral signs of life in the atmospheres of exoplanets is a premier goal of modern astronomy. While future large-aperture space telescopes with coronagraphs may enable the direct imaging and spectroscopy of Earth-like exoplanets orbiting stars similar to our sun (see e.g. Ref.~\cite{Ruane2018_JATIS}), the next-generation of ground-based telescopes with adaptive optics will focus on planets in the habitable zone of cooler M~dwarf stars, such as the known planets Proxima~Centauri~b~\cite{ProxCen} and Ross~128~b~\cite{Bonfils2017}. The number of planets expected to be detected and spectroscopically characterized with both space-based and ground-based facilities is, however, limited by the inner working angles of their respective high contrast imaging systems. Improving sensitivity at smaller angular separations provides access to many more potential targets whose planet-to-star flux ratios are favorable due to their close proximity to their host star and increases the maximum wavelength at which they can be observed.

\begin{figure*}[t!]
    \centering
    \includegraphics[width=0.9\linewidth]{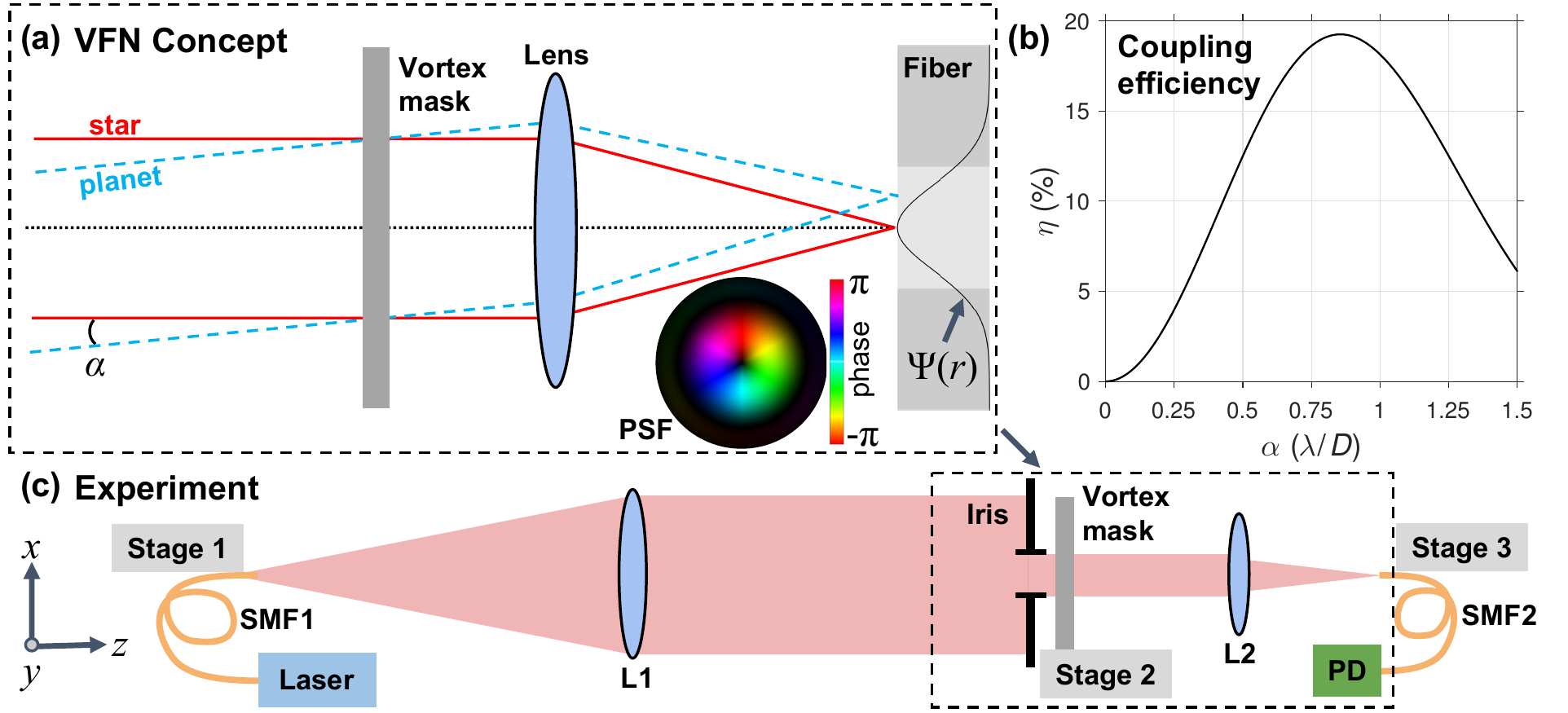}
    \caption{(a)~Schematic of a vortex fiber nuller (VFN). Light from the star (red rays) and planet (blue rays) passes through a vortex mask with complex transmittance $\exp(\pm il\theta)$. The image of the star is aligned to a single-mode fiber on the optical axis with fundamental mode, $\Psi(r)$, whereas the planet image is slightly off-axis. Each has a characteristically donut-shaped point spread function (PSF) and phase proportional to the azimuthal angle; the inset shows the simulated phase and PSF. (b) The coupling efficiency, $\eta$, evaluated for any point at angular separation $\alpha$, assuming the star is at $\alpha=0$. (c)~Diagram of the prototype VFN demonstrated here. Light from a fiber-coupled laser is launched by SMF1 and collimated by lens L1 (focal length $f=200$~mm) to evenly illuminate an iris defining the pupil. The beam passes through a liquid crystal vector vortex mask directly behind the iris and is focused by lens L2 ($f=11$~mm) onto SMF2. We measure the coupled power using photodiode PD. Stages 1 and 3 have five-axes while stage 2 is a two-axis stage. Linear actuators provide computer control of the $x$ and $y$ directions for the vortex mask and $x$, $y$, and $z$ for SMF2.}
    \label{fig:diagram}
\end{figure*}

We present an experimental demonstration of an optical system known as a vortex fiber nuller (VFN)~\cite{Ruane2018_VFN} that allows for the spectral characterization of exoplanets at angular separations less than the Rayleigh criterion; i.e. $<1.22~\lambda/D$, where $\lambda$ is the wavelength and $D$ is the telescope diameter. Figure~\ref{fig:diagram}a illustrates a VFN with a vortex phase mask~\cite{Beijersbergen1994} placed in a pupil plane to impart a phase pattern of the form $\exp(\pm il\theta)$ as in Ref.~\cite{Swartzlander2001}, where $l$ is an integer known as the charge. This prevents the starlight  from coupling into a single-mode fiber (SMF) which is actively aligned with the star's geometric image in the focal plane. The stellar point spread function (PSF) is rejected by the SMF because its complex field is orthogonal to the fiber's fundamental mode. For an arbitrary point source, the fraction of light that couples into the fiber as a function of its angular separation from the optical axis, $\alpha$, is
\begin{equation}
    \eta(\alpha) = \frac{\left| \int  \Psi(\mathbf{r}) f(\mathbf{r}; \alpha) dA\right|^2}{\int \left| \Psi(\mathbf{r}) \right|^2 dA \int \left| f(\mathbf{r};\alpha) \right|^2 dA},
    \label{eqn:overlap}
\end{equation}
where $\Psi(\mathbf{r})$ is the fiber mode and $f(\mathbf{r};\alpha)$ is the field in the final image plane~\cite{Shaklan1988}. $dA$ is the differential area and $\mathbf{r}=(r,\theta)$ are polar coordinates in the $(x,y)$ plane. For common SMFs, the fundamental mode can be approximated as a Gaussian with the functional form $\Psi(r)=\exp[-(2r/D_f)^2]$, where $D_f$ is the mode field diameter. Any stellar field of the form $f(\mathbf{r};0)=f_r(r)\exp(\pm il\theta)$ leads to $\eta(0)=\eta_\mathrm{star}=0$. However, light from a point source (e.g. a planet) at an angular separation $\alpha$ will couple into the fiber with the efficiency shown in Fig.~\ref{fig:diagram}b. The maximum theoretical coupling efficiency in this arrangement, $\eta=19\%$, is achieved at $\alpha=0.86~\lambda/D$ with $D_f=1.4~\lambda F\#$, where $F\#$ is the focal ratio of the lens. In practice, the single-mode fiber is routed to a spectrograph which is used to separate the starlight as well as measure and analyze spectral signatures in the planet light~\cite{Wang2017}.% \ul{ Any leaked starlight that couples into the fiber can be spectrally separated from the planet signal due to the Doppler shift between the two. Therefore, as shown in} \cite{Ruane2018_VFN} \ul{, this seemingly narrow $>10\%$ coupling region enables the characterization of earth-size planets in reflected light, such as Ross 128b, at separations and wavelengths that are currently inaccessible by other methods.}

Vortex fiber nulling is akin to traditional nulling interferometry~\cite{Bracewell1978,Haguenauer2006}, but makes use of the full telescope aperture. The key advantages are (1)~the extremely small inner working angle, (2)~the rotational symmetry of the coupling function, allowing for spectral follow up of planets when the azimuthal orientation of their orbits are uncertain, and (3)~the theoretically wavelength-independent nulling of starlight across astronomically relevant bandwidths ($\Delta\lambda/\lambda\approx0.2$). Here, we demonstrate the predicted nulling effect in the laboratory using a prototype system designed for monochromatic light.

\section{Experimental Setup}
Figure~\ref{fig:diagram}c shows a diagram of the experimental optical layout. Light from a SMF-coupled laser diode operating at $\lambda=635$~nm is collimated by a 200~mm focal length lens (L1) and evenly illuminates a 3.6~mm diameter iris. 
A charge $l{=}1$, liquid crystal vector vortex mask (Thorlabs WPV10L-633) immediately after the iris applies the desired phase pattern to the incoming beam in the pupil plane. Computer-controlled linear actuators (Zaber X-NA08A25) on stage~2 provide fine transverse alignment ($<1~\mu$m repeatability) of the vortex mask in the $x$ and $y$ directions.

The vortex mask is a half wave plate with a spatially-variant fast axis angle, $\chi=l\theta/2$. The transmitted complex field is $E_{R,L}=\exp(\pm i 2\chi)E_{L,R}$, where $E_R$ and $E_L$ are the right and left circularly polarized components, respectively~\cite{Marrucci2006}. The two output components have conjugate vortex phases of the form $\exp(\pm il \theta)$ and a polarization state that is orthogonal to the input. The mask is optimized to provide the half wave retardance, and thus the vortex phase, at a single wavelength.%, $\lambda=633$~nm.\ul{ The 2~nm difference in central wavelength between the vortex and laser limits the null to $\sim10^{-7}$, a negligible effect relative to the degradation from wavefront errors in the system.}

An 11~mm focal length aspheric lens (L2) focuses the beam onto the detection fiber (SMF2) centered on the optical axis and connected to a variable-gain silicon photodiode (PD; Femto OE-200-SI). The path lengths between the iris, vortex, and L2 are 5 and 35mm respectively, set by the size of their mounts. SMF1 and SMF2 are both SM600 fibers with $D_f=$3.6-5.3~$\mu$m and a measured single-mode cutoff wavelength of <550~nm.
Each is fixed to a 5-axis stage (stages 1 and 3; degrees of freedom: $x$, $y$, $z$, tip, and tilt). Computer-controlled piezo actuators (Thorlabs PE4) on stage~3 position SMF2 to an accuracy of 10~nm (15~$\mu$m travel) in the $x$, $y$, and $z$ directions.

We used a Shack-Hartmann wavefront sensor (ImagineOptic HASO4-Broadband) to minimize the static aberrations during alignment. The total wavefront error, through both lenses and with the vortex mask in the beam but offset so as not to create a singularity in the phase, was 7.3~nm~RMS ($\sim\lambda/100$).

We determined the maximum coupling efficiency by translating the vortex mask such that the beam passed through a region far from the phase singularity and then co-aligning SMF1 and SMF2. In this configuration, the coupling efficiency was 56\% which is in close agreement with the theoretical value of 57\% expected from our $F\#$=3.1 system assuming an optimal $F\#$ of 5 provided that the SMF2 core diameter is 4.45~$\mu m$ (manufacturer specification is 3.6-5.3~$\mu m$). We attribute 0.4\% of the coupling losses to the measured wavefront error. Improved coupling efficiency is possible by matching the $F\#$ with the ideal value given the measured fiber core diameter and using a custom lens.
% \ul{We determined the maximum coupling efficiency by moving the vortex mask such that the beam passed through a region far from the phase singularity and then co-aligning SMF1 and SMF2. In this configuration, the coupling efficiency was 56\% which is in close agreement with the theoretical value of 57\% expected from our $F\#$=3.1 system given the poorly constrained focal ratio required for SMF2, $F\#$=4.1-6. We attribute 0.4\% of the coupling losses to the measured wavefront error. The peak coupling could be increased by measuring the mode-field diameter of the fiber to determine its ideal F\#. This would also improve the planet coupling efficiency reported in the Results.}
% %In this configuration, the coupling efficiency was 56\% rather than the theoretical maximum of 82\%. We attribute 0.4\% of the coupling \ul{loss to wavefront error and the remainder to a mismatch between the ideal focal ratio for SMF2 ($F\#$=4.1-6.0) and the actual one for our system ($F\#$=3.1).} Indeed, for a $F\#$=3.1 beam, the optimum theoretical coupling efficiency is 57\%, which is in good agreement with the measured value. 

\section{Procedure}

The objective of the experimental procedure was (1)~to demonstrate that light from an on-axis, unresolved source is rejected by SMF2 and (2)~to show that light from an off-axis source couples into SMF2 with the predicted efficiency. Since the system PSF is shift-invariant, we can simulate a planet by translating SMF1 or SMF2 in the $(x,y)$ plane; both are equivalent barring a magnification factor. Thus, for convenience and to maintain low wavefront error, we opted to measure the coupling efficiency as a function of the position of SMF2 using highly-accurate piezo actuators. 

To ensure that aberrations were minimized, we removed stage~3 and SMF2 and took images of the PSF with a CMOS detector (Thorlabs DCC1545M) with and without the vortex mask. Then, with stage~3 and SMF2 back in place, we performed several two-dimensional raster scans of SMF2 in a 7$\times$7~$\mu$m square, adjusting the position of the vortex mask between each scan until we minimized the coupling for the simulated star, $\eta_\mathrm{star}$.   

At each SMF2 position, we checked that the power measured by the PD was above the predetermined noise floor and then averaged 100 measurements before moving SMF2 to the next location. Once each full 2D scan was completed, we inserted a calibrated power meter (Thorlabs PM100D, S120C) in front of SMF2 and measured the total power in the beam to normalize the signal at the output. Finally, we determined the bias signal of the PD by blocking the light source and subtracted the bias from our measurements. To obtain $\eta$, we normalized the measured power by the total power accounting for the transmission of SMF2, including the Fresnel reflections at both ends (3.46\% per facet) as well as propagation losses (0.34\% per meter).

\begin{figure}[t]
    \centering
    \includegraphics[width=\linewidth]{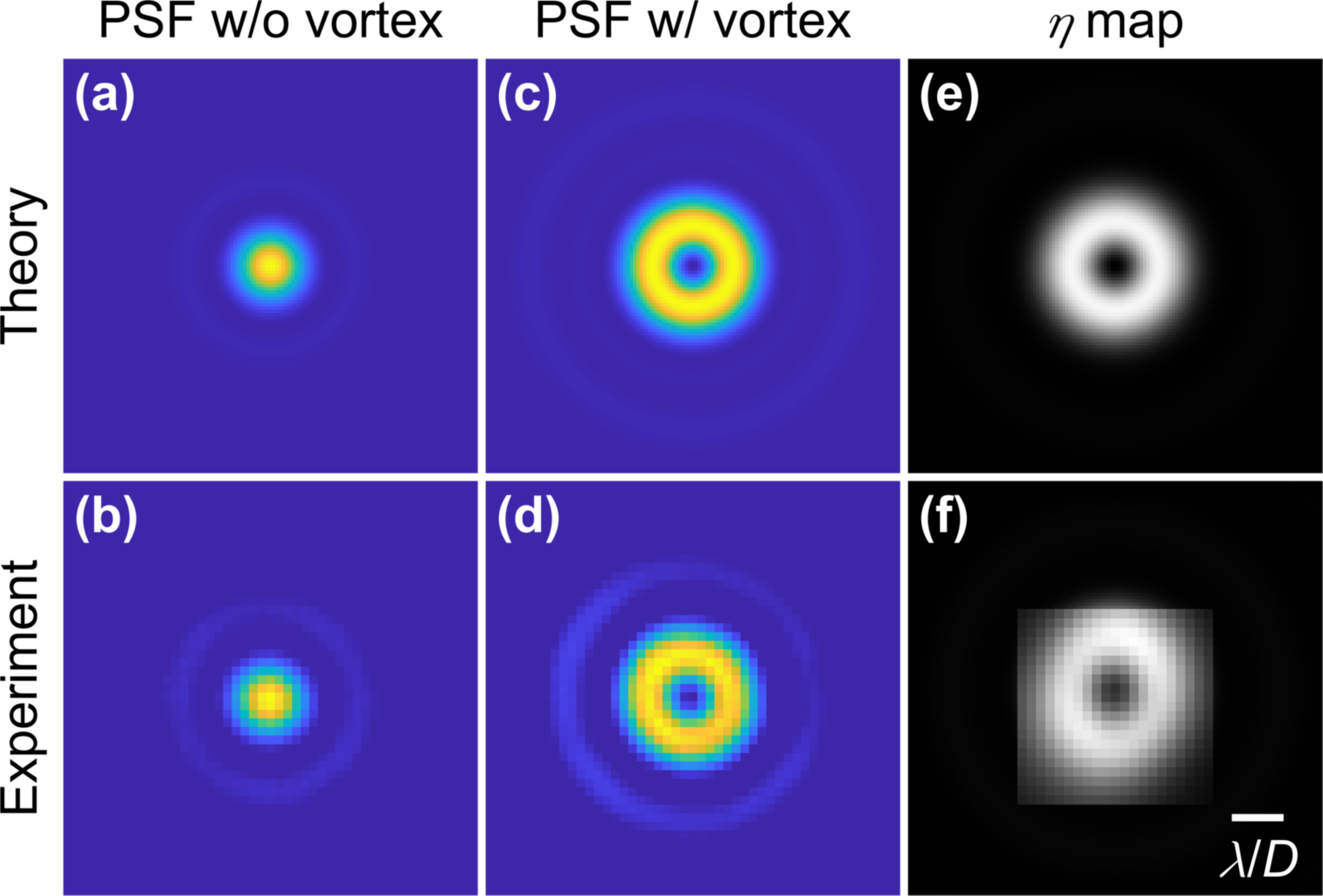}
    \caption{PSFs (a)-(b) without and (c)-(d) with the vortex mask centered on the pupil. (e)-(f) Coupling efficiency, or "$\eta$ map", as a function of the position of SMF2, equivalent to the source, with respect to the optical axis ($0.35~\mu$m step size). The theoretical predictions match well with our experimental results. The PSF images and coupling map have six samples per $\lambda F\#$.
    }
    \label{fig:psfs_and_etamaps}
\end{figure}

\section{Results}
Figure~\ref{fig:psfs_and_etamaps} shows images of the PSF at the plane of SMF2 (Figure~\ref{fig:psfs_and_etamaps}a-d) and the two-dimensional coupling maps obtained by scanning SMF2 in the $(x,y)$ plane. The PSF images with the beam passing through the edge of the vortex mask (Fig.~\ref{fig:psfs_and_etamaps}b) resembles an Airy pattern (Fig.~\ref{fig:psfs_and_etamaps}a) validating that the collimated beam evenly illuminates the iris. The PSF with the vortex centered on the iris (Fig.~\ref{fig:psfs_and_etamaps}d) appears annular in shape as predicted (Fig.~\ref{fig:psfs_and_etamaps}c). The ideal coupling efficiency as a function of the 2D position of SMF2 (see Fig.~\ref{fig:psfs_and_etamaps}e) is a donut shape with the radial profile in Fig.~\ref{fig:diagram}b. The measured coupling map (Fig.~\ref{fig:psfs_and_etamaps}f) shows a very similar shape except for a slight vertical elongation likely owing to imperfect calibration of the piezo actuator gains. 

\begin{figure}[t]
    \includegraphics[width=0.87\linewidth]{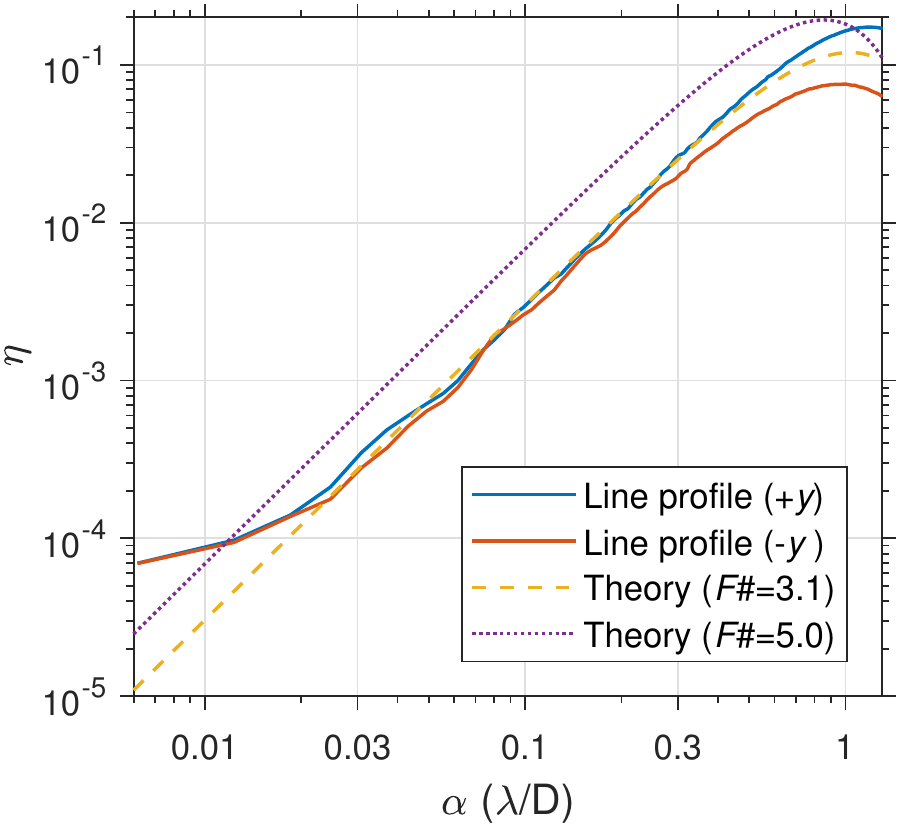}
    \caption{Linear scans of SMF2 in the $\pm y$ directions with a 12~nm step size. The best achieved null $\eta_\mathrm{star}=\eta(0)=6\times10^{-5}$, is likely limited by static aberrations. 
    Our measurements match well with the theoretical line profiles for our $F\#=3.1$ system but $F\#=5.0$ is required to achieve the ideal performance.}
    \label{fig:lineprofiles}
\end{figure}

In addition to the 2D coupling map in Fig.~\ref{fig:psfs_and_etamaps}f, we took two fine linear scans starting at the deepest null found and moving radially outward in the $\pm y$ directions in 12~nm steps (see Fig.~\ref{fig:lineprofiles}). The deepest null measured, with SMF1 and SMF2 on the optical axis, was $\eta_\mathrm{star}=6\times10^{-5}$~$(0.006\%)$. The maximum coupling efficiency for the line scans, corresponding to the peak planet coupling, was $\eta=8\%$ and $15\%$ for the $\pm y$ directions, respectively, revealing an asymmetry in the coupling map. We compare our line scan measurements with the theoretical line profiles for an ideal system ($F\#$=5.0) and our setup ($F\#$=3.1). The laboratory-measured line profiles are in close agreement with our theoretical predication showing a minor departure at $\alpha\lesssim0.02~\lambda/D$ and $\alpha\gtrsim0.3~\lambda/D$. We expect that optimizing the focal ratio would improve the theoretical maximum coupling efficiency from 12\% up to a maximum of $19\%$ and the peak would shift from $\alpha=\lambda/D$ to $\alpha=0.86~\lambda/D$.  

\section{Discussion}
Though the experimental PSFs (Fig.~\ref{fig:psfs_and_etamaps}b,d) are in close agreement with theory (Fig.~\ref{fig:psfs_and_etamaps}a,c), the diffraction rings are slightly more pronounced than expected. This is likely due to spherical aberration unseen by our wavefront sensor, which we used in a diverging beam after L2's focus and therefore discarded all power terms.

Using our wavefront measurements, taken at the focus with the vortex mask offset such that the phase singularity was outside of the iris, we estimate the minimum possible $\eta_\mathrm{star}$ by numerically computing Eqn. \ref{eqn:overlap}. As in Ref. \cite{Ruane2018_VFN}, we determined the sensitivity of the null to low-order Zernike aberrations, $Z_n^m$. For each Zernike mode, we fit the response function $\eta_\mathrm{star} = (b\omega)^{2}$, where $b$ is the sensitivity coefficient and $\omega$ is the RMS wavefront error in units of waves. By orthogonality, the sensitivity coefficient is zero for all Zernike modes $Z_n^m$ where $m\ne\pm1$. Also, the precision of the piezo actuators mitigates the tip and tilt errors. Thus, our experimental nulls are likely limited by coma aberrations. Table~\ref{tab:aberrations} lists the measured wavefront error in the coma modes ($m=\pm1$) as well as their corresponding $b$ value and predicted contribution to $\eta_\mathrm{star}$. Taking the linear combination of the error contributions predicts a minimum of $\eta_\mathrm{star}=5.0\times10^{-5}$, which is in good agreement with the minimum measured value, $\eta_\mathrm{star}=6.0\times10^{-5}$.

\begin{table}[]
    \centering
    \caption{Measured wavefront error, $\omega$, in the coma Zernike modes, $Z_n^{\pm1}$, and expected stellar leakage, $\eta_\mathrm{star}=(b\omega)^2$, for ($0^{\circ}$,$90^{\circ}$) aberrations, where $b$ is the aberration sensitivity coefficient. All other modes have $b=0$.}
    \begin{tabular}{lccc}
        \hline
         Modes & $\omega$ (waves RMS) & $b$ & $\eta_s$\\
        \hline
        $Z_3^{\pm1}$ & (-1.7,-2.8)$\times10^{-3}$ & 2.15 & (1.3,3.6)$\times10^{-5}$ \\
        $Z_5^{\pm1}$ & (0.1,-0.2)$\times10^{-3}$ & 1.12 & (1.3,5.0)$\times10^{-8}$ \\
        $Z_7^{\pm1}$ & (-0.7,0.0)$\times10^{-3}$ & 0.67 & (2.2,0.0)$\times10^{-7}$ \\
        \hline
        \multicolumn{1}{r}{\textbf{Total}} &  &  & \textbf{$5.0\times10^{-5}$} \\ 
        \hline
    \end{tabular}
\label{tab:aberrations}
\end{table}

The line scans presented in Fig.~\ref{fig:lineprofiles} reveal an asymmetry in peak coupling efficiency around the donut. In that case, to achieve the smallest value of $\eta_\mathrm{star}$, we scanned both the position of the vortex mask in addition to SMF2. Using numerical simulations, we find that, in the presence of coma aberrations, the optimal null occurs when the vortex mask is slightly off-center causing the coupling map to become asymmetric.  
This implies that, when observing an exoplanet with a known orbit around its host star, it may be possible to deliberately misalign the vortex in the pupil to create an asymmetric coupling map which preferentially couples more light at the location of the planet. Maximizing the throughput for the planet is important as the integration time scales as the inverse square of the planet's coupling efficiency, $\eta^{-2}$, in the stellar photon noise limited regime \cite{Ruane2018_VFN}. 

The pupil shape has little influence on the VFN's performance~\cite{Ruane2018_VFN}; the results presented here are also valid for non-circular, obstructed, and segmented apertures. In fact, the pupil created by the iris used in these experiments was only quasi-circular, with $10$ flat edges. Future experiments will use a pupil mask that mimics the boundaries of an actual telescope pupil. 

We have demonstrated the VFN concept in monochromatic light using a simple, inexpensive optical system. However, exoplanet spectroscopy requires similar starlight suppression levels in polychromatic light  ($\Delta\lambda/\lambda\approx0.2$). We plan to build a polychromatic testbed to demonstrate this using off-axis parabolic mirrors instead of lenses, a carefully matched $F\#$, and broadband vortex masks optimized for the optical and infrared. Vortex masks that apply the same phase pattern as a function of wavelength have been demonstrated using polarization dependent, or "vector," methods: liquid crystals~\cite{Mawet2009}, sub-wavelength gratings~\cite{Mawet2005b}, and photonic crystal structures~\cite{Murakami2013}. Achromatic scalar masks are also possible~\cite{Swartzlander2006}.

Furthermore, we plan to integrate a polychromatic VFN into an adaptive optics system with a deformable mirror, similar to previous fiber injection instruments tested by our team~\cite{Mawet2017_HDCII}, which will allow us to develop wavefront sensing and control techniques to maintain the null in the presence of realistic wavefront errors and flux levels. Ultimately, our goal is to integrate a VFN module into the Keck Planet Imager and Characterizer at the W.M. Keck Observatory~\cite{Mawet2017_KPIC} to open up the possibility of characterizing the reflected light spectrum of giant exoplanets whose properties have so far only been inferred from stellar radial velocity measurements. This will pave the way to characterizing smaller, potentially habitable planets with future large-aperture telescopes.

\section{Conclusions}
We have demonstrated the VFN concept in a laboratory for the first time. Using a prototype system designed for monochromatic light, we demonstrated a stellar coupling fraction of $\eta_\mathrm{star}=6.0\times10^{-5}$ and peak planet coupling efficiencies of $\eta$=8-15\% at an angular separation of $\alpha\approx\lambda/D$. These results match the expected performance for our F\#=3.1 setup and thereby validate the model described in Ref.~\cite{Ruane2018_VFN}. As such, we predict that using the ideal $F\#$ for the fiber and minimizing wavefront errors will yield $\eta=19\%$ at $0.86~\lambda/D$ in all azimuthal directions. We have also identified a clear pathway to achieving similar performance in polychromatic light and to developing the wavefront sensing and control techniques needed for on-sky operation. 

The VFN concept is a promising approach for reducing the stellar photon noise that otherwise inhibits the characterization of exoplanets whose angular separations are within the inner working angle of conventional coronagraphs. We expect that this technique will open the possibility to measure the reflected light spectrum of exoplanets inferred from stellar radial velocity measurements and thereby allow for the detailed characterization of their atmospheres for the first time. Several confirmed planets are at angular separations of $\alpha\lesssim\lambda/D$ in the infrared, which is currently too close to the star to characterize by other means, but lie within the collecting region of a VFN. We envisage that vortex fiber nulling will significantly increase the number of exoplanets, from rocky worlds to gas giants, characterized by current and future ground- and space-based telescopes. 

\medskip
\noindent\textbf{Funding.} National Science Foundation (NSF) (AST-1602444);

\noindent\textbf{Acknowledgments.} Part of this work was supported by the Jet Propulsion Laboratory, California Institute of Technology, under contract with NASA.

% Bibliography
\bibliography{sample}
\bibliographyfullrefs{sample}

\end{document}